\documentclass[%
reprint,
 amsmath,amssymb,
 aps,
pra,
superscriptaddress,
]{revtex4-1}

\usepackage[english]{babel}
\usepackage{amsmath,amssymb}
\usepackage{setspace}
\usepackage{hyperref} 
\usepackage[utf8x]{inputenc}
\usepackage{caption}
\usepackage{leftidx}
\usepackage{graphicx}
\usepackage{epstopdf}
\usepackage{dcolumn}
\usepackage{bm}
\usepackage{float}

\begin{document}

\title{Influence of the asymmetric excited state decay on coherent population trapping: atom $\times$ quantum dot}

\preprint{APS/123-QED}

\author{H. S. Borges}
\author{M. H. Oliveira}
\author{C. J. Villas-B\^{o}as}
\affiliation{Departamento de F\'{i}sica, Universidade Federal de S\~{a}o Carlos, P.O. Box 676, 13565-905, S\~{a}o Carlos, S\~{a}o Paulo, Brazil\\}



\begin{abstract}
Electromagnetically induced transparency (EIT) is an optical phenomenon which allows a drastic modification of the optical properties of an atomic system by applying a control field.  It has been largely studied in the last decades and nowadays we can find a huge number of experimental and theoretical related studies. Recently a similar phenomenon was  also shown in quantum dot molecules (QDM) , where  the control field is replaced by the tunneling  rate between quantum dots. Our results show that in the EIT regime, the optical properties of QDM and the atomic system are identical. However, here we show that in the strong probe field regime, i.e.,  "coherent population trapping" (CPT)  regime, it appears a strong discrepancy on the optical properties of both systems. We show that the origin of such difference relies on the different decay rates of the excited state of the two systems, implying in a strong difference on their higher order nonlinear susceptibilities. Finally, we investigate the optical response of atom/QDM strongly coupled to a cavity mode.  In particular, the QDM-cavity system has the advantage of allowing a better narrowing of the width of the dark state resonance in the CPT regime when compared with atom-cavity system.
\end{abstract}

\flushbottom
\maketitle
%
%
\thispagestyle{empty}



It is well known that quantum interference between different excitation paths can modify the optical response of a system, giving rise for example to the suppression of absorption of the incident light when the interference between these channels is destructive. Optical nonlinear effects, such as electromagnetically induced transparency (EIT) \cite{Boller1991, Marangos05}, result in the suppression of a weak probe field absorption in a narrow spectral window accompanied by an enhancement of its nonlinear susceptibility and an abrupt change of the refraction index \cite{Imamoglu90, Lukin01}. The electromagnetically induced transparency phenomenon is associated with another process named coherent population trapping (CPT)\cite{Matisov93, Gammon08}, which is characterized by a dark state written as a coherent superposition of the ground states of the system. In this context, quantum interference and coherent effects have been widely investigated and demonstrated in various three-level systems in $\Lambda$-configuration, which can be modelled in atomic \cite{Marangos05, Villas_Boas10} and semiconductor systems, for example coupled quantum dots (QDM - Quantum Dot Molecule)  \cite{Imamoglu12}. Nowadays we can find a huge number of theoretical and experimental studies on EIT, presenting applications such as slowing down of light pulses \cite{Hau1999}, quantum memories in atomic ensembles \cite{Liu2001,Phillips2001} or in optical cavities \cite{Specht2011}, cooling down trapped atoms \cite{Morigi2000, Roos2000, Lechner2016}, among many others.

Placing the atom or the QDM inside an optical cavity, their couplings with the probe field is replaced by coupling with a cavity mode (strenght $g$), which is driven by the probe laser. For practical proposes the interaction of single emitters (atom or quantum dot) with optical cavity modes has a fundamental role, enabling a significant enhancement of the light-matter coupling and the increase of the efficiency of photon collection. In the atom-cavity system the EIT effect can be observed through the transmission spectrum of an incident laser field (probe laser), which will be entirely transmitted at resonance if the control field is present \cite{Zhu11}. In this context it is important remind that the linewidth of the transmitted field depends on the ratio between the squares of the control field Rabi frequency $\Omega_C$, and atom-cavity coupling $g$ \cite{Villas_Boas10}, which is valid when there is no decay or dephasing on the atomic ground states and when the system is in the EIT regime.

In this work we investigate the difference between the optical response of two distinct three-level systems in $\Lambda$-configuration, i.e., a single atom and QDM. In both systems there are two ground states coupled to the same excited state -- see Fig. \ref{fig:0} (a) and (b). In the free-space the couplings are mediated by the probe laser and by the control field for atomic system. For the QDM, the control laser beam is replaced by the electron tunneling between the quantum dots (with tunneling rate $T_e$), which can be controlled by an external electric field \cite{Gammon06b}. This effect is known as tunneling induced transparency (TIT) due to the critical role of tunneling in the appearance of transparency in QDM's \cite{Yuan2006, Borges12} and allows for applications similar to those we find in atomic systems, e.g., slow down of light pulses \cite{Yuan2006} or cavity linewidth narrowing \cite{Peng2011,Peng2014}.

Our results show a notable difference in the optical response of these systems when the Rabi frequency of the probe laser is comparable or larger than the Rabi frequency of the control field (tunneling rate), i.e., $\Omega_P\gtrsim\Omega_C(T_e)$. Besides, we note a very appreciable difference in the transmission spectrum of the cavity when we have atom or QDM coupled to it and when we are in the limit $\Omega_C \rightarrow 0$ or $T_e \rightarrow 0$, respectively. In those limits and when all the fields are on resonance, while the transmission of the atom-cavity system reaches an empty cavity situation, the QDM-cavity presents an extremely narrow transmission peak. Thus, here we show that QDM works out more efficiently to induce cavity-linewidth narrowing \cite{Lukin1998,Wang2000,Peng2011,Peng2014} than atoms. Our analysis shows that these features occurs due to nonlinear effects, which become more pronounced in the weak control field limit (CPT regime). We show that the crucial parameter which allows for the enhancement of the nonlinear effects in EIT/CPT processes is the asymmetric decay rate of the excited state. 

\section*{Results}

\subsection*{EIT and TIT in Free Space}
\label{sec:2}

\begin{figure}[h]
\centering
\includegraphics[width=0.8\linewidth]{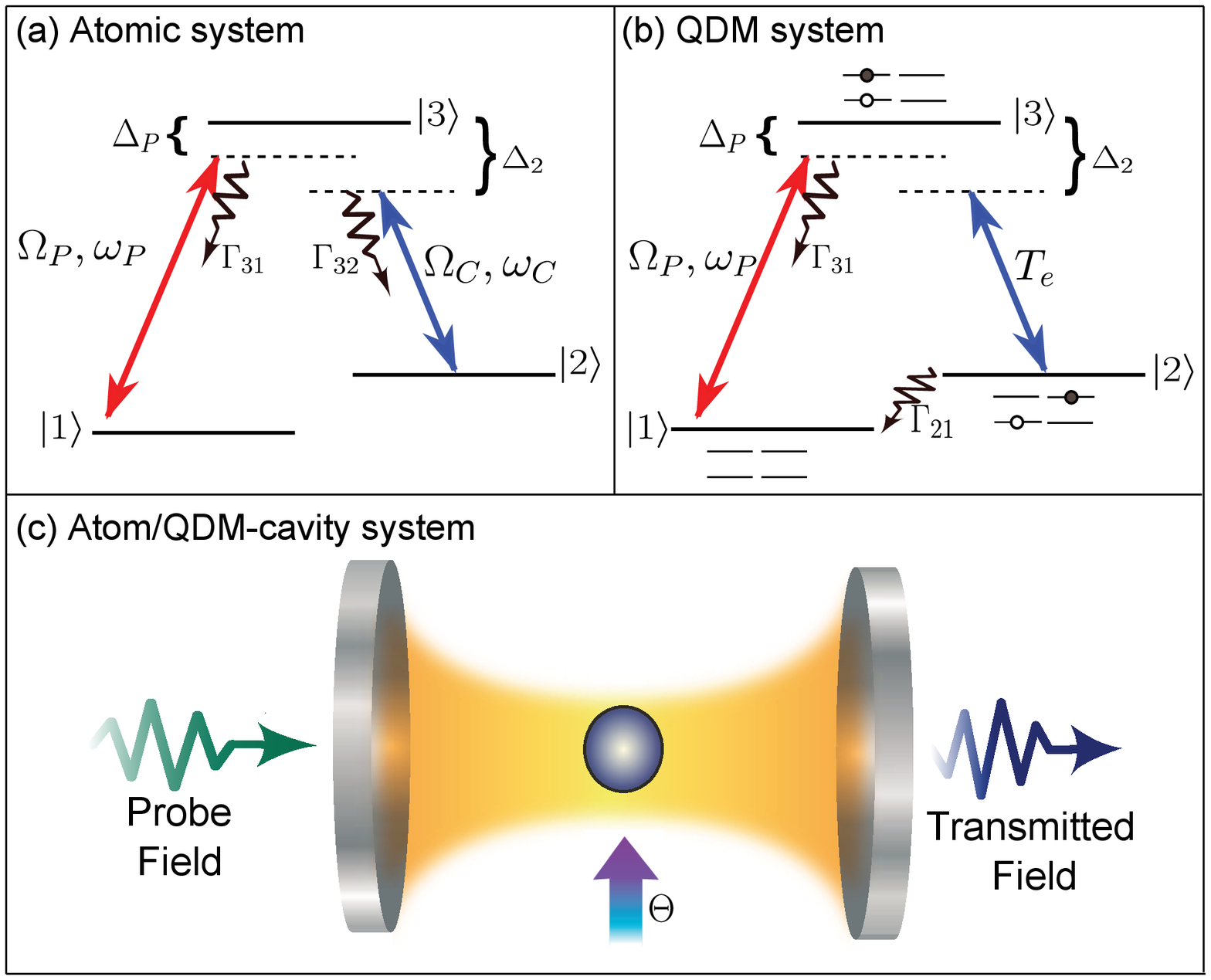}
\caption{\textbf{Schematic representation of the system.} Diagram of levels: (a) for the atomic and (b) QDM systems. In (c) we show an atom/QDM-cavity system which can be used to verify  the nonlinear effects predicted in the present work and to investigate the cavity-linewidth narrowing. The parameters which appear here are explained in the text.}
\label{fig:0}
\end{figure} 

Both the atom and the QDM can be modelled by the same general Hamiltonian, i.e, a three-level system in a $\Lambda$-configuration with two ground states, $|1\rangle$ and $|2\rangle$, and an excited one $|3\rangle$, as shown in Fig. \ref{fig:0}. The levels $|1\rangle $ and $ |3\rangle $ (transition frequency $\omega_{31}$) are coupled by a probe field with Rabi frequency $\Omega_P$ while the levels $|2\rangle $ and $ |3\rangle $ (transition frequency $\omega_{32}$) are either coupled by a classical control field (frequency $\omega_C$) with Rabi frequency $\Omega_C$ or by a tunnelling process with tunnelling rate $T_e$ for the atomic or QDM system, respectively. In the QDM system, the energy levels correspond to the excitonic states (electron-hole pair held together by their attractive Coulomb interaction), where the ground state $|1\rangle$ is the QDM without any excitation, and the states $|2\rangle$ and $|3\rangle$ are the exciton states with indirect and direct character, respectively \cite{Jose_Maria04}. Considering the rotating wave approximation, the Hamiltonian that describes both atomic and QDM systems (without temporal dependency), can be written in the interaction picture as ($\hbar=1$)
\begin{equation}
\label{eq:1}
H_{I}=\Delta_P \sigma _{11}- \left( \frac{\Omega_P}{2} \sigma_{31}+\frac{\Theta}{2} \sigma _{32}+ h.c. \right) \text{,}  
\end{equation}
being $\Delta_P = \omega_P - \omega_{31}$ the detuning between the $|1\rangle \leftrightarrow |3 \rangle$ transition and the probe field ($\omega_P$) frequencies. The atomic or QDM operators are represented by $\sigma _{kl}=\left\vert k\right\rangle \left\langle l\right\vert $ ($k,l=1,2,3$). $\Theta = \Omega_C$ or $T_e$, for the atom or for the QDM, respectively, and $h.c.$ stands for Hermitian conjugate. The dissipation of the system can be included in the dynamics of system through the master equation

\begin{equation}
\frac{d\rho }{dt}= -i[H_{I},\rho ] +\sum_{i=1}^{3} \sum^3_{\substack{k=2 \\(k \geqslant i) }} \frac{\Gamma _{ki}}{2}(2\sigma _{ik}\rho \sigma _{ki}-\sigma_{ki}\sigma _{ik}\rho -\rho \sigma _{ki}\sigma _{ik}),
\label{eq:2}
\end{equation}%
being $\Gamma _{ki}$ ($k\ne i$) the decay rate of the level $|k \rangle$ to level $|i\rangle$ and $\Gamma_{ii}$ the dephasing rate of the level $|i\rangle$. Here we can point out the main difference between the atomic and QDM systems: while usually we find $\Gamma _{32} \approx \Gamma_{31} \neq 0$ for atoms, $\Gamma_{31} \neq 0$ and $\Gamma _{32} =0 $ for QDM. This difference will introduce a strong modification on the nonlinear behaviour of atoms and QDM as we discuss below. Of course, one could find an atomic system in $\Lambda$-configuration where $\Gamma_{31} \gg \Gamma_{32}$ and then its optical response would be similar to that of the QDM we are analysing here.

\begin{figure}[h]
\centering
\includegraphics[width= 1.0\linewidth]{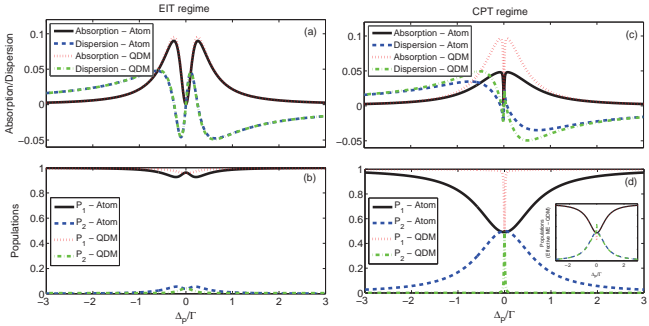}
\caption{\textbf{Optical response of the atomic and quantum dot molecule systems in two different regimes of parameters, i.e., EIT and CPT.} In all plots we have considered a total decay $\Gamma$ of the excited state $|3 \rangle$. We have fixed $\Omega_P = 0.1 \Gamma$ and $\Theta = 0.5 \Gamma$  for the EIT regime, panels (a) and (b), and  $\Omega_P = 0.1 \Gamma$ and $\Theta = 0.1 \Gamma$  for the CPT regime, panels (c) and (d). In (a) and (c) we plot the Absorption ($Im\langle \sigma_{13} \rangle$) and the Dispersion ($Re\langle \sigma_{13} \rangle$) of the systems and in panels (b) and (d) we plot the populations of the ground states $|1 \rangle$ ($P_1$) and $|2 \rangle$ ($P_2$). In all plots the solid black and blue dashed lines refer to the case $\Gamma_{31} = \Gamma_{32} = 0.5 \Gamma$ (atomic system) while the red dotted and green dashed-dotted lines refer to $\Gamma_{31} = \Gamma$ and $\Gamma_{32} = 0$ (Quantum Dot Molecule system). Inset: populations of the QDM system obtained via the effective master equation, i.e., taking into account the effective decay from level $|1\rangle$ to $|2\rangle$.}
\label{fig:1}
\end{figure}

\begin{figure}[h]
\centering
\includegraphics[width=1.0\linewidth]{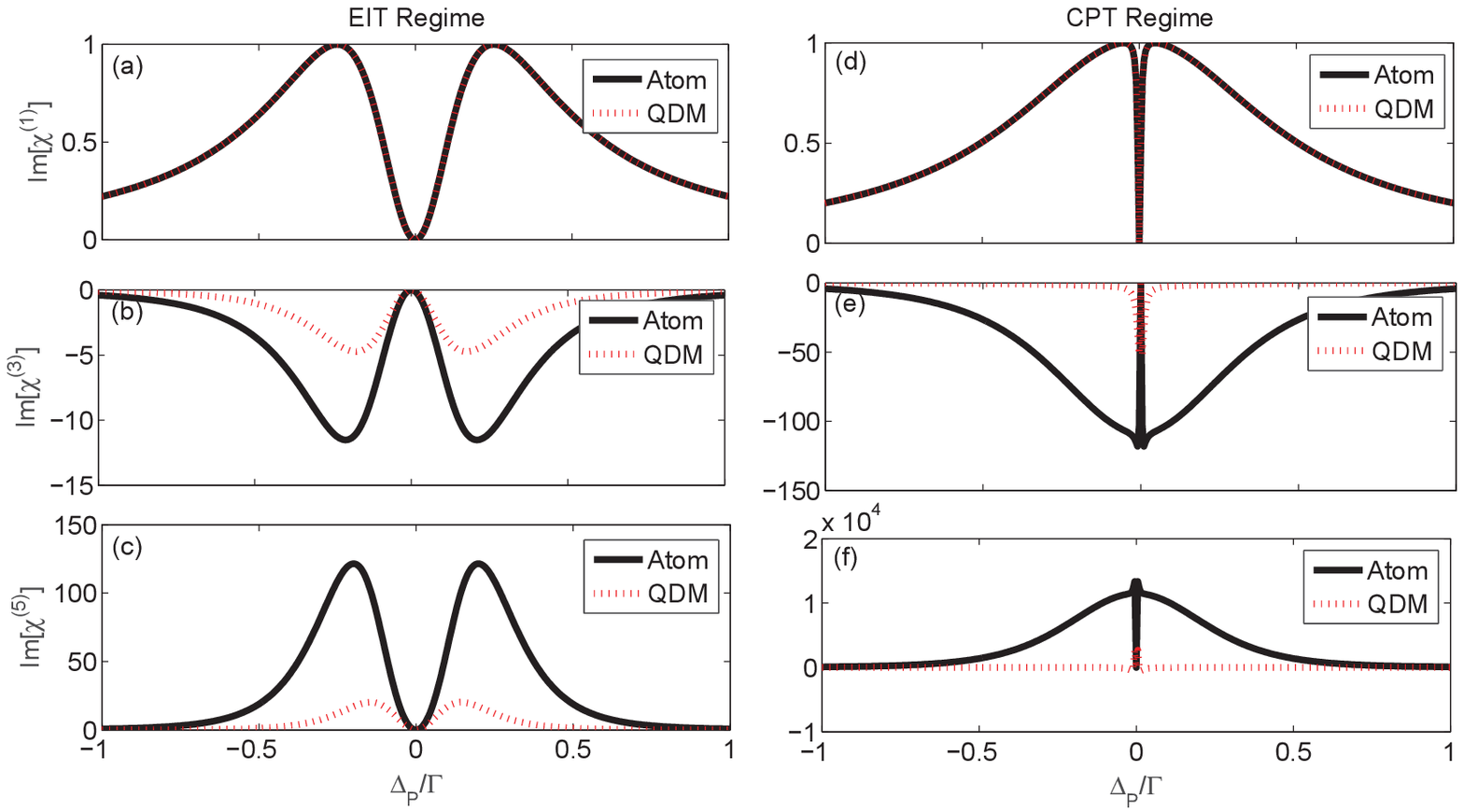}
\includegraphics[width=1.0\linewidth]{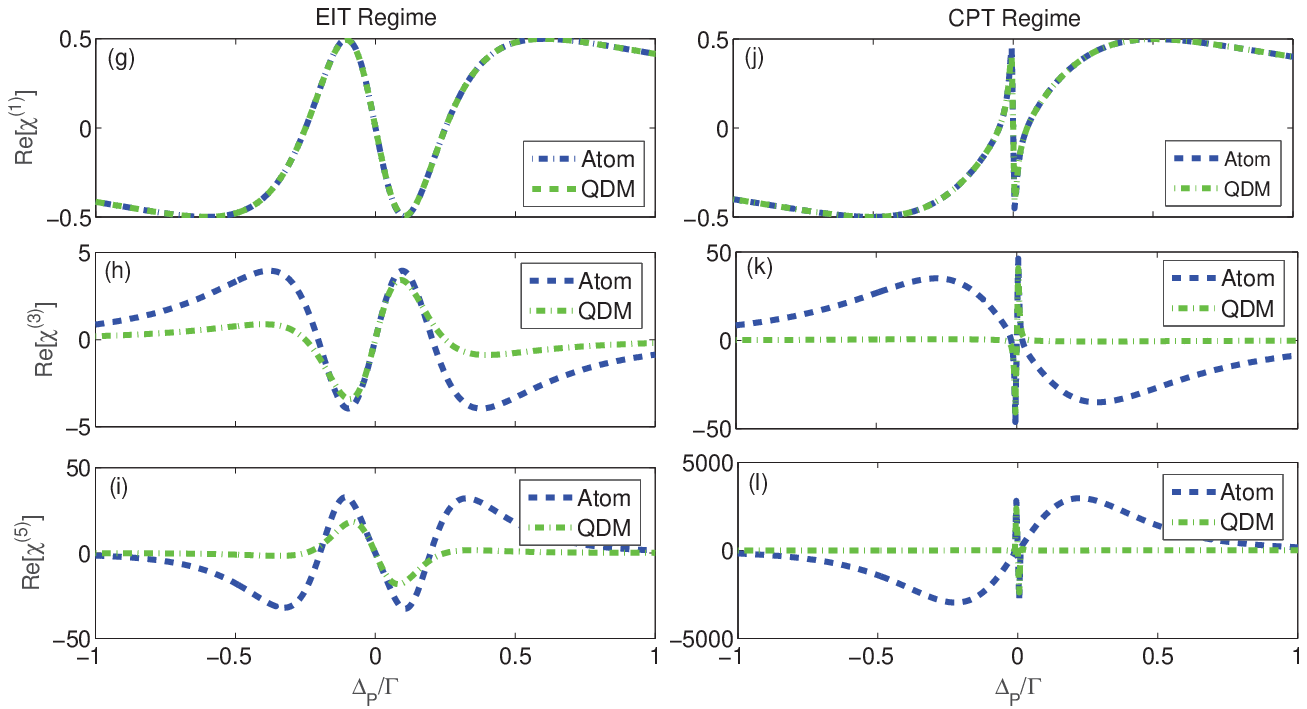}
\caption{\textbf{Linear and nonlinear optical susceptibilities.} In panels (a) to (f) [and from (g) to (l)] we plot the imaginary [real] part of $\chi^{(n)}$ ($n = 1,3,5$) for the atomic/quantum dot molecule systems in two different regimes of parameters, i.e., EIT (left panels) and CPT (right panels), considering the same set of parameters as in Fig. \ref{fig:1}.}
\label{fig:2}
\end{figure}

We can easily obtain the steady state solution for the master equation (\ref{eq:2}). Considering different regimes of control (tunnelling process) and probe fields we immediately see expressive differences of the optical response for atomic and QDM systems, here quantified by \textit{Absorption} and \textit{Dispersion} defined, apart from a scaling factor, as $Im(\langle \sigma_{13} \rangle)$ and $Real(\langle \sigma_{13} \rangle)$, respectively. In all plots of Fig. \ref{fig:1} we have considered  $\Gamma_{31} = \Gamma_{32} = 0.5\Gamma$ for the atoms and $\Gamma_{31} = \Gamma$ and $\Gamma_{32} = 0$ for the QDM, so that in both cases the total decay rate of the excited state $|3 \rangle$ is $\Gamma$. Throughout this work we have neglected the other decoherence and dissipation rates, i.e., $\Gamma_{22}=\Gamma_{33}= \Gamma_{21}=0$, since they are much smaller than the decay rates of the excited state and also because they destroy dark state of the system, thus making difficult the comparison between the optical responses of the atomic and QDM systems. We note that, when $\Omega_P \ll \Theta =(\Omega_C,T_e)$, i.e., in the regime of parameters known as EIT regime (or even in the Autler-Townes regime), the optical response of the system is independent on the decay rate of the excited state $|3\rangle$ to the ground state $|2\rangle$, as we can see in Fig. \ref{fig:1} (a). However, for $\Omega_P \gtrsim \Theta $, i.e., in the regime known as CPT (coherent population trapping) we can see a strong difference between the atomic and QDM optical responses, as shown in Fig. \ref{fig:1} (c). The width of the transparency window is the same for both systems. However, the QDM presents higher absorption in the region around the transparency window. The origin of such difference is on the decay rate $\Gamma_{32}$, which is non-null for the atomic system but null for QDM. In Fig. \ref{fig:1} (b) and (d) we plot the populations of the states $|1 \rangle$ ($P_1$) and $|2\rangle$ ($P_2$) for atomic and QDM systems (the population of the excited state $|3 \rangle$ is close to zero in all cases), for EIT and CPT regimes, respectively. Again, there is an expressive difference between the behaviour of the populations of the atom and of the QDM in the CPT regime. In particular, for QDM the populations $P_1$ and $P_2$ are modified only for resonant probe field, i.e., for $\Delta_P =0$. The origin of such difference between QDM and atoms is an effective decay from level $|1 \rangle$ to $|2 \rangle$, described by the effective Lindbladian $ \frac{\Gamma _{12}^{eff}}{2}(2\sigma _{21}\rho \sigma _{12}-\sigma_{11}\rho -\rho \sigma _{11})$, which arises only for non-null $\Gamma_{32}$, being $\Gamma_{12}^{eff}=P_3\Gamma_{32}$. In the limit of $\Theta \rightarrow 0$ and for $\Gamma_{32}=0$ we would have a two-level system driven by a probe field (Rabi frequency $\Omega_P$) and with a total decay rate of the excited state given by $\Gamma$. In this case, the steady population of the excited state is \cite{Walls2008} $P_3 = 1/2 \left( 1+\langle \sigma_{33} - \sigma_{11} \rangle\right) = \frac{|\Omega_P|^2}{\Gamma^2 + 4 \Delta_P^2 + 2 |\Omega_P|^2}$. Introducing the effective decay rate from level $|1 \rangle$ to $|2 \rangle$ in the dynamics of the QDM we can reproduce almost perfectly the populations of the atomic system, as we see in the inset of Fig.\ref{fig:2}(d) (only exactly on the dark state resonance we see a discordance, which becomes negligible when we decrease even more the value of $\Theta$).  

To understand how the decay rate $\Gamma_{32}$ affects the optical response of our system, it is instructive to analyse the optical susceptibility of the system and its linear and nonlinear components. The polarization density of the system is given by \cite{Boyd2008} $\vec P = \chi^{(1)} \vec E_P + \chi^{(3)} \vec E ^3_P + \chi^{(5)} \vec E ^5_p +...$, being $\chi^{(n)}$ the $n_{th}$-order susceptibility of the medium. On the other hand, by solving the master equation (\ref{eq:2}) we are able to obtain the polarization density in terms of the density matrix elements $\rho_{ij}(t)$ ($i,j = 1,2,3$) \cite{Marangos05}. By imposing $d\rho/dt = 0$ we can get the steady state solution of the master equation (\ref{eq:2}), which allows us to obtain the $n_{th}$-order optical susceptibility \cite{Marangos05} which can be written as (without a scaling factor):
\begin{equation}
\chi^{(1)} = \frac{2\Delta_P}{2\Delta_P\left(2\Delta_P - i \Gamma\right) - \Theta^2}, 
\label{eq:3}
\end{equation}

\begin{equation}
\chi^{(3)} = \frac{\left(\frac{\Gamma^2\Gamma_{32}}{\Theta^2}-i\frac{\Gamma\Gamma_{32}}{2\Delta_P}+3\Gamma-\Gamma_{31}+\frac{\Gamma_{31}\Theta^2}{2\Delta_P^2}\right)}{\Gamma_{31}\left(\frac{\Theta^2}{2\Delta_P}-i\Gamma-2\Delta_P\right)\left(\frac{\Theta^2}{2\Delta_P}+i\Gamma-2\Delta_P\right)^2}.
\label{eq:5}
\end{equation}
Here, $\Gamma_3=\Gamma_{31}+\Gamma_{32}= \Gamma$. The expressions above are valid for any values of $\Omega_P$ and $\Theta$ so that these results are valid in both EIT and CPT regimes. From Eq. (\ref{eq:3}) we see that $\chi^{(1)}$ depends only on the total decay rate of the excited state $|3\rangle$ ($\Gamma_3$), being not important the value of the decay rate $\Gamma_{32}$ alone. This behaviour is shown in Fig. \ref{fig:2} (a) and (d), and (g) and (j) as well, where we plot the $Im(\chi^{(1)})$ and $Re(\chi^{(1)})$ versus $\Delta_P/\Gamma$, respectively. In all plots we consider the total decay rate $\Gamma_3=\Gamma$ and the same parameters as those used in Fig.\ref{fig:1}. On the other hand, the higher order optical susceptibilities ($\chi^{(3)}$ and $\chi^{(5)}$) present a strong dependency on $\Gamma_{32}$, which is evidenced in the others panels of Fig. \ref{fig:2} where we plot $Im(\chi^{(n)})$ and $Re(\chi^{(n)})$ ($n=3,5$) versus $\Delta_P/\Gamma$. Through these results we see that $\Gamma_{32}$ enhances the nonlinear processes, mainly in the limit of $\Theta \lesssim \Omega_P$. Although there is a strong difference of the optical response between the systems ($\Gamma_{32} =0 $ and  $\Gamma_{32} \neq 0$) in the region around the dark state resonance, exactly at $\Delta_P = 0$ the systems are equivalent: both predict null absorption ($Im(\chi^{(3)})$) and the same slope of the dispersion ($Real(\chi^{(3)})$), which result in the same group velocity of light pulses when interacting with either atoms or QDM's.

Expanding the nonlinear optical susceptibility $\chi_{Im}^{(3)}$ in a power series of $\Theta$ also helps us to mathematically see how $\Gamma_{32}$ affects the optical response of the system in the limit of weak control field:

\begin{equation}
\label{eq:8}
\begin{split}
\chi_{Im}^{(3)}=&-\frac{2\Gamma^2\left(\Gamma^2+2\Gamma\Gamma_{32}+4\Delta_P^2\right)}{\Gamma_{31}\left(\Gamma^2+4\Delta_P^2\right)^3}-\frac{\Gamma^3\Gamma_{32}}{\Gamma_{31}\Theta^2\left(\Gamma^2+4\Delta_P^2\right)}\\ &+A\Theta^2,
\end{split}
\end{equation}
where $A=\frac{4\Gamma\left(\frac{\Gamma^4\left(\Gamma_{32}-2\Gamma_{31}\right)}{16\Delta_P^2}-\Gamma^3\left(3\Gamma+2\Gamma_{32}\right)-\Delta_P^2\left(9\Gamma+\Gamma_{31}\right)\right)}{\Gamma_{31}\left(\Gamma^2+4\Delta_P^2\right)^4}$.

As can be seen in the expression (\ref{eq:8}), the first term is independent of $\Theta$ and the second one has $\Theta^2$ at denominator and is zero if $\Gamma_{32}=0$. Such fact indicates that this term is one of the main responsible for the enhancement of the non linear effects of the optical susceptibility associated to $\Gamma_{32}$. In fact, the last term in this expression goes to zero in the limit of $\Theta \rightarrow 0$, while the second one becomes more and more relevant in this limit. This is exactly what we observe in Fig. \ref{fig:2}: out of the transparency window, decreasing $\Theta$ from $0.5\Gamma$ (EIT regime) to $0.1\Gamma$ (CPT regime), the maximum (in absolute values) value of $\chi^{(3)}$ has an increasing of the order of 10 times, when $\Gamma_{32} \ne 0$, but its variation is negligible when $\Gamma_{32} = 0$.

\subsection*{Cavity EIT and TIT}
\label{sec:3}

\begin{figure}[h]
\centering
\includegraphics[width=1.0\linewidth]{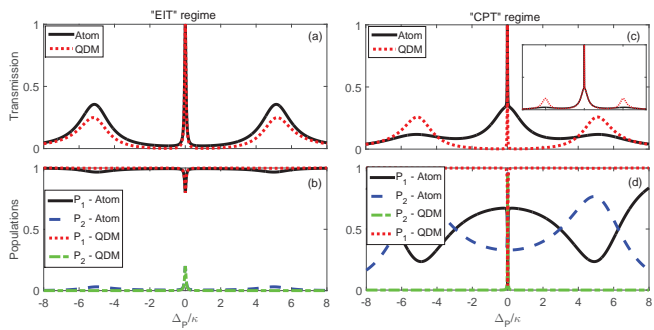}
\caption{\textbf{Transmission spectrum and populations.} Painels (a) and (c) (top) show the normalized cavity transmission and panels (b) and (d) (bottom) show the atomic and QDM populations, both as a function of the detuning $\Delta_P/\kappa$. The parameters used here are $\kappa = \Gamma$, $g=5\kappa$, $\varepsilon = \sqrt{0.01}\kappa$, and $\Theta = 1.0 \kappa$ for the EIT (panels (a) and (b), on the left) and $\Theta = 0.1 \kappa$ for the CPT regimes (panels (c) and (d), on the right), and $\Gamma_{31} = \Gamma_{32} = 0.5\Gamma$ for the atomic system and $\Gamma_{31} = \Gamma$ and $\Gamma_{32} = 0$ for the QDM one. For small values of the control field Rabi frequency or Tunnelling rate ($\Theta \rightarrow 0$), the non null $\Gamma_{32}$ leads the population of the system to the ground state $|2 \rangle$, which is not coupled to the cavity mode, thus increasing the transmission (empty cavity situation). This is equivalent to an effective decay from level $|1\rangle$ to $|2\rangle$ -- see inset of panel (c), where we have included the effective decay in the QDM-cavity dynamics.}
\label{fig:4}
\end{figure}

Making use of strong QDM/atom-cavity couplings, the effects predicted above could be experimentally investigated at a level of single atom/QDM and used, e.g., in applications such as cavity-linewidth narrowing \cite{Lukin1998,Wang2000,Peng2011,Peng2014}. A single atom or a single QDM coupled to a cavity mode can be described by the following master equation \cite{Oliveira16}
\begin{equation}
\begin{split}
\frac{d\rho }{dt}=& -i[H_{cav},\rho ]+\frac{\kappa}{2} (2a\rho a^{\dagger }-a^{\dagger}a\rho -\rho a^{\dagger }a)\\ &+\sum_{i=1}^{3} \sum^3_{\substack{k=2 \\(k \geqslant i) }} \frac{\Gamma _{ki}}{2}(2\sigma _{ik}\rho \sigma _{ki}-\sigma_{ki}\sigma _{ik}\rho -\rho \sigma _{ki}\sigma _{ik}),
\end{split}
\label{eq:9}
\end{equation}%
being $\kappa$ the total decay rate of the intensity of the cavity field and $a$ and $a^{\dagger}$ the annihilation and creation operators associated to the internal cavity mode, respectively. Considering for simplicity the $|1\rangle \leftrightarrow |3\rangle$ ($|2\rangle \leftrightarrow |3\rangle$) transition resonant to the cavity mode (control field/tunneling), the  Hamiltonian $H_{cav}$ given in the interaction picture and without time-dependency is ($\hbar=1$): 
\begin{equation}
\label{eq:10}
H_{cav}=\Delta_P (\sigma _{11}-a^{\dagger }a)+{\left(\frac{\varepsilon}{2} a+ ga\sigma_{31}+\frac{\Theta}{2} \sigma _{32}+ h.c. \right)}\text{,} 
\end{equation}
where $\Delta_P = \omega_P - \omega$ represents the detuning between the cavity mode ($\omega$) and the probe field ($\omega_P$) frequencies. The atom/QDM-cavity system is probed by a weak laser represented by a driving field of strength $\varepsilon$. Finally, $g$ is the atom/QDM-cavity coupling. Again, we can solve the master equation in the steady state regime ($d\rho/dt = 0$) and then we can derive the main optical properties of the system. We do this numerically by properly truncating the Hilbert space of the cavity mode and then using QuTip algorithms \cite{Johansson2012}. Going back to Fig. \ref{fig:1} (b), we note that the populations of the system does not present a substantial difference when we have null (QDM) or non-null (atom) $\Gamma_{32}$ for the EIT regime of parameters ($\Theta \gg \Omega_P$). However, the populations in the CPT regime ($\Omega_P\gtrsim\Theta$) are strongly dependent on $\Gamma_{32}$, as we see in Fig. \ref{fig:1} (d): for $\Gamma_{32} = 0$ (QDM) $P_1 \simeq 1$ and $P_2 \simeq 0$ for all values of $\Delta_P$ except for $\Delta_P \simeq 0$, i.e., exactly on the dark state resonance. This means that the linewidth of the dark state for the QDM case can be narrower than the atomic case. When we place the atom/QDM inside a cavity, its transmission will be strongly affected by this difference of populations (for $\Gamma_{32}$ null and non-null cases). In Fig. \ref{fig:4} we plot the normalized cavity transmission ($\langle n \rangle /\max{ \langle n \rangle} $, with $ n =  \langle a^{\dagger} a \rangle$) of the atom/QDM-cavity systems, either in the EIT or in the CPT regimes. In both cases we have considered $\kappa = \Gamma$. Again, for the atom we have considered $\Gamma_{31} = \Gamma_{32} = 0.5\Gamma$ and for the QDM, $\Gamma_{31} = \Gamma$ and $\Gamma_{32} = 0$. The other parameters are $\kappa = \Gamma$, $g=5\kappa$, $\varepsilon = \sqrt{0.01}\kappa$, and $\Theta = 1.0 \kappa$ for the EIT and $\Theta = 0.1 \kappa$ for the CPT regimes.

As it happens in free space, when $\Theta \rightarrow 0$, there will be an effective decay from level $|1 \rangle$ to $|2 \rangle$, whose effective decay rate is $P_3 \Gamma_{32}$, which modifies considerably the dynamics of the system when $\Gamma_{32} \neq 0$ (atoms). This is indeed the case, as we can see in the inset of Fig. \ref{fig:4}(c), where we plot the transmission of the QDM-cavity system with an additional effective decay from level $|1\rangle$ to $|2\rangle$ (decay rate $ P_3 \Gamma_3/2$). When $\Theta \rightarrow 0$ we have a two-level atom interacting with a driven cavity mode. In this case, and considering the probe field close to resonance (around the EIT peak) we have \cite{Walls2008} $\langle \sigma_{33} -\sigma_{11} \rangle =  -n_0/(n+n_0)$, with $n_{0} = \Gamma_{3} /8 g^2$ and $n = |\alpha|^2 = \left| \frac{\varepsilon/\kappa(n+n_0)}{n+n_0+2Cn_0} \right|^2 $, being $C = 2g^2/\kappa \Gamma_{3}$ the Cooperativity. In the limit of weak driving field ($\varepsilon \rightarrow 0$, which implies $n \ll n_0$) and large Cooperativity ($C \gg 1$) we obtain $P_3  = 1/2(1+\langle \sigma_{33} -\sigma_{11}\rangle ) \approx |\varepsilon/2g|^2$, resulting in an effective decay rate from level $|1 \rangle$ to $|2 \rangle$ given by $\Gamma^{eff}_{12} = |\varepsilon/2g|^2 \Gamma_{32}$. Including this dissipation channel in the master equation of the QDM-cavity system we recover approximately the atom-cavity dynamics (around the EIT peak), as we see in the inset of Fig. \ref{fig:4}(c).

\begin{figure}[h]
\centering
\includegraphics[width=1.0\linewidth]{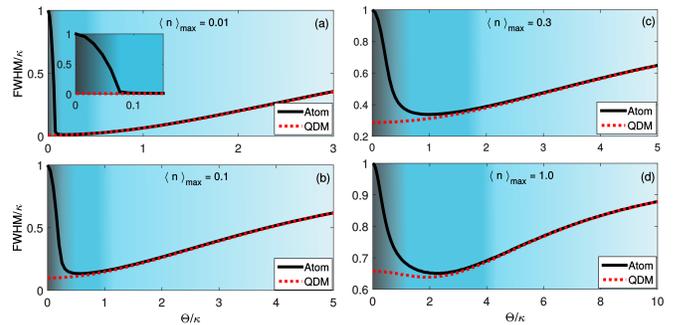}
\caption{\textbf{Full Width at Half Maximum ($FWHM$) of the dark state resonance.} $FWHM$ for the atomic (black solid line) and QDM (red dotted line) systems as a function of $\Theta$ (=$\Omega_C$ or $T_e$). Here assumed $\Gamma_2 = \Gamma_{21} = 10^{-3}\Gamma$ and the other parameters are the same as those used in Fig. \ref{fig:4}, except for $\varepsilon$, which  implies different maximum average number of photons inside the cavity ($\langle n \rangle_{\max} = |\varepsilon/\kappa|^2$): (a)  $\varepsilon = \sqrt{0.01}\kappa$, (b) $\varepsilon = \sqrt{0.1}\kappa$, (c) $\varepsilon = \sqrt{0.3}\kappa$, and (d) $\varepsilon = \sqrt{1.0}\kappa$. In some cases, the minimum $FWHM$ for QDM becomes $10\%$ narrower than the minimum $FWHM$ for atomic system.}
\label{fig:5}
\end{figure}

As we see in Fig. \ref{fig:4}(c), when we decrease $\Omega_C$ ($T_e$), i.e., in the CPT regime, the transmission around $\Delta_P = 0$ increases for the atomic system while for the QDM system it remains non-null only exactly on the dark state resonance, i.e., at $\Delta_P = 0$. This happens because the population of the ground state $|2\rangle$ increases due to the presence of the decay channel associated to $\Gamma_{32}$ as $\Omega_C \rightarrow 0$, making the atom-cavity system transparent to the probe field (empty cavity situation). On the other hand, in the QDM system, i.e., for $\Gamma_{32} = 0$, in the limit $T_e \rightarrow 0$ we end up with a perfect two-level system, which presents null transmission at $\Delta_P \approx 0$ and strong coupling regime ($g\gg \kappa,\Gamma_{31}$). In this way, here we have an interesting difference between atomic and QDM behaviors: while the atomic system reaches an empty cavity transmission profile when $\Omega_C \rightarrow 0 $, the QDM system presents a very narrow transmission peak when $T_e \rightarrow 0 $. In Fig. \ref{fig:5} we plot the full width at half maximum ($FWHM$) of the dark state resonance for the atomic and QDM systems as a function of $\Theta$ (= $\Omega_C$ or $T_e$). We note that the $FWHM$ reaches a minimum value for the atomic system and then starts increasing as the population of the ground state $|2 \rangle$ increases (for $\Omega_C$ $\rightarrow 0 $). (We have started with very small values of $\Theta$, i.e., $\Theta_{min}=0.001\kappa$ since for $\Theta = 0$ the QDM reduces to a two-level system and then it is not possible to define $FWHM$.) The $FWHM$ is also dependent on the strength of the probe field ($\varepsilon$), as we see in \ref{fig:5}(a) to (d). For very small mean number of photons inside the cavity, the probability of excitation of the atom/QDM is also very small. For instance, for maximum average number of photons $0.01$, Fig. \ref{fig:5}(a), the minimum $FWHM$ for atom is very close to that for QDM. However, the difference between atom and QDM increases when we increase the average number of photons. Increasing $\varepsilon$ we get a higher probability of having two (or more) photons inside the cavity. But, as we are considering only a single atom/QDM interacting with the cavity mode, it can absorb only a single photon from the probe field. In this way, the minimum $FWHM$ also depends on the number of atoms/QDMs inside the cavity \cite{Borges2017}. Thus, considering the application in cavity-linewidth narrowing based on cavity EIT as proposed in \cite{Lukin1998} and experimentally verified by H. Wang \textit{et al.} \cite{Wang2000}, our results show that QDM allows to reach $FWHM$ narrower than atomic systems in the limit of $\Theta \rightarrow 0$, thus being more attractive for this kind of application. As we see in Fig. \ref{fig:5}(b), the minimum $FWHM$ for atom becomes $35\%$ broader than the minimum $FWHM$ for QDM system. For stronger values of $g$ the minimum $FWHM$ for atomic and QDM systems becomes closer.



\section*{Conclusion}
\label{sec:4}
We have investigated the influence of the asymmetric decay of the excited state of three-level systems in $\Lambda$-configuration on the EIT and CPT processes. When the decay rate $\Gamma_{31} \gg \Gamma_{32}$, which can be found in Quantum Dot Molecule (where $\Gamma_{32}=0$), the nonlinear susceptibilities $\chi^{(3)}$ or $\chi^{(5)}$ is much smaller (in absolute values) than in the case where $\Gamma_{32} \approx \Gamma_{31}$, which is usually the case for atomic systems. Thus, the high nonlinear effects present in EIT or CPT experiments are strongly related to the decay channel $\Gamma_{32}$, which is associated to the transition coupled by the control field. On the other hand, non-null $\Gamma_{32}$ implies in high population of the system in the state $|2\rangle$ in the limit of $\Theta$ $\ll \Omega_P $. When this system is placed inside a cavity, this results in an empty cavity situation, increasing its transmission. On the order hand, for $\Gamma_{32}=0$ we end up with a perfect two-level system resonantly coupled to a cavity mode, which presents null transmission when the probe field is resonant with the cavity mode and in the strong atom/QDM-cavity coupling. Thus, only exactly on resonance we have the dark state, implying a very narrow transmission peak for the QDM when compared with atomic systems. In this way, we hope this work can be useful for studies on nonlinear effects in EIT or CPT processes and future applications in cavity-linewidth narrowing.

\section*{Acknowledgements}
C. J. V.-B. and H. S. Borges acknowledge support from CNPq and FAPESP (Proc. 2013/04162-5 and 2014/12740-1), and the Brazilian National Institute for
Science and Technology of Quantum Information (INCT-IQ). M.H. O. acknowledges support from CNPq.

\section*{Author contributions statement}
C.J.V.B. conceived the project, produced the codes for the simulation and wrote the manuscript. H.S.B. obtained the analytical results, helped with the simulations and edited the figures. M.H.O. helped with the simulations and helped to get the effective dynamics of the systems. All the authors contributed equally with all discussions and explanations of the results, reviewed the text and gave the final approval for the publication of this paper.     

\section*{Additional information}
\textbf{Competing financial interests:} The authors declare no competing financial interests. 

%

\end{document}